\pdfoutput=1
\documentclass[9pt,a4paper,twoside,twocolumn]{article}
\usepackage{amsmath,amsfonts,amssymb}
\usepackage{array}
\usepackage[american]{babel}
\usepackage{balance}
\usepackage[format=plain,justification=justified,singlelinecheck=false,font={stretch=1.125,small,sf},labelfont=bf,labelsep=space]{caption}
\usepackage{extsizes}
\usepackage[T1]{fontenc}
\usepackage[utf8]{inputenc}
\usepackage{lastpage}
\usepackage{microtype}
\usepackage{multirow}
\usepackage[numbers,square,super,sort&compress,comma]{natbib}
\usepackage[left=1.5cm, right=1.5cm, top=1.785cm, bottom=2.0cm]{geometry}
\usepackage{sectsty}
\usepackage{float}
\usepackage{fancyhdr}
\usepackage{fnpos}
\usepackage{graphicx}
\usepackage[linkdoi,usedoi,usetitle]{rsc}
\usepackage{setspace}
\usepackage{times,mathptmx}
\usepackage[usenames,dvipsnames]{xcolor}
\usepackage[compact]{titlesec}
\usepackage[textsize=footnotesize,obeyFinal]{todonotes} 
\usepackage{hyperref}
\usepackage{textcomp}

\relax
\RequirePackage{ifthen}
\if@ams\RequirePackage{mathtools}\fi
\RequirePackage{textcomp}
\RequirePackage{xspace}

\newcommand*{\oricosta}{\ensuremath{\left<\cos\!\alpha_\text{12,2D}\right>}\xspace}
\newcommand*{\costa}{\ensuremath{\left<\cos^2\!\alpha_\text{12,2D}\right>}\xspace}

\newcommand*{\eg}{e.\,g.}%
\newcommand*{\ie}{i.\,e.}%
\newcommand*{\ordsim}{\mathord{\sim}}
\newcommand*{\celsius}[1]{\ensuremath{#1\,^\circ{}\text{C}}}%
\newcommand*{\degree}[1]{\ensuremath{#1\,^\circ}}%
\newcommand*{\us}{\ensuremath{\text{\textmu{s}}}}
\newcommand*{\subautoref}[2]{\autoref{#1}\,#2}

\definecolor{cream}{RGB}{222,217,201}

\makeatletter%
\def\@fnsymbol#1{\ensuremath{\ifcase#1\or \dagger\or \ddagger\or \mathsection\or \mathparagraph\or
      \|\or **\or \dagger\dagger \or \ddagger\ddagger \else\@ctrerr\fi}}%
\makeatother%

\begin{document}
\pagestyle{fancy}
\thispagestyle{plain}
\fancypagestyle{plain}{}

\makeFNbottom
\makeatletter
\renewcommand\LARGE{\@setfontsize\LARGE{15pt}{17}}
\renewcommand\Large{\@setfontsize\Large{12pt}{14}}
\renewcommand\large{\@setfontsize\large{10pt}{12}}
\renewcommand\footnotesize{\@setfontsize\footnotesize{7pt}{10}}
\makeatother
\renewcommand{\thefootnote}{\fnsymbol{footnote}}
\renewcommand\footnoterule{\vspace*{1pt}%
\color{cream}\hrule width 3.5in height 0.4pt \color{black}\vspace*{5pt}}
\setcounter{secnumdepth}{5}
\makeatletter
\renewcommand\@biblabel[1]{#1}
\renewcommand\@makefntext[1]{\noindent\makebox[0pt][r]{\@thefnmark\,}#1}
\makeatother
\renewcommand{\figurename}{\small{Fig.}~}
\sectionfont{\sffamily\Large}
\subsectionfont{\normalsize}
\subsubsectionfont{\bf}
\setstretch{1.125} 
\setlength{\skip\footins}{0.8cm}
\setlength{\footnotesep}{0.25cm}
\setlength{\jot}{10pt}
\titlespacing*{\section}{0pt}{4pt}{4pt}
\titlespacing*{\subsection}{0pt}{15pt}{1pt}

\fancyfoot{}
\fancyfoot[RO]{\footnotesize{\sffamily{1--\pageref{LastPage}~\textbar~\thepage}}}
\fancyfoot[LE]{\footnotesize{\sffamily{\thepage~\textbar~1--\pageref{LastPage}}}}
\fancyhead{}
\renewcommand{\headrulewidth}{0pt}
\renewcommand{\footrulewidth}{0pt}
\setlength{\arrayrulewidth}{1pt}
\setlength{\columnsep}{6.5mm}
\setlength\bibsep{1pt}

\makeatletter
\newlength{\figrulesep}
\setlength{\figrulesep}{0.5\textfloatsep}
\newcommand{\topfigrule}{\vspace*{-1pt}\noindent{\color{cream}\rule[-\figrulesep]{\columnwidth}{1.5pt}} }
\newcommand{\botfigrule}{\vspace*{-2pt}\noindent{\color{cream}\rule[\figrulesep]{\columnwidth}{1.5pt}} }
\newcommand{\dblfigrule}{\vspace*{-1pt}\noindent{\color{cream}\rule[-\figrulesep]{\textwidth}{1.5pt}} }
\makeatother

\twocolumn[
\begin{@twocolumnfalse}
   \sffamily
   \LARGE{\textbf{Photophysics of indole upon x-ray absorption}} \\
   \\
   \large{Thomas Kierspel,$\!^{a,b,c}$ %
      C\'edric Bomme,$\!^{d}$ %
      Michele Di Fraia,$\!^{a,b,e}$ %
      Joss Wiese,$\!^{a,f}$ %
      Denis Anielski,$\!^d$ %
      Sadia Bari,$\!^{d,g}$ %
      Rebecca Boll,$\!^{d,g}$ %
      Benjamin Erk,$\!^{d}$ %
      Jens S. Kienitz,$\!^{a,b,c}$ %
      Nele L.\ M.\ Müller,$\!^{a,d}$ %
      Daniel Rolles,$\!^{d,h}$ %
      Jens Viefhaus,$\!^{d}$ %
      Sebastian Trippel,$\!^{a,b}\,^{\ast}$ and %
      Jochen Küpper$\,^{a,b,c,f}$} \\
   \\
   \noindent\normalsize{A photofragmentation study of gas-phase indole (C$_8$H$_7$N) upon
      single-photon ionization at a photon energy of~420~eV is presented. Indole was primarily
      inner-shell ionized at its nitrogen and carbon $1s$ orbitals. Electrons and ions were measured
      in coincidence by means of velocity map imaging.
      The angular relationship between ionic fragments is discussed along with the possibility to use
      the angle-resolved coincidence detection to perform experiments on molecules that are strongly
      oriented in their recoil-frame. The coincident measurement of electrons and ions revealed
      fragmentation-pathway-dependent electron spectra, linking the structural fragmentation dynamics
      to different electronic excitations. Evidence for photoelectron-impact self-ionization was observed.
   } \\
   \begin{center}
      \begin{minipage}{0.9\linewidth}
         \emph{David W.\ Pratt originally initiated our investigations into the photophysics of
            indole and this paper is dedicated to him on the occasion of his 80\/$^\text{th}$
            birthday.} \\
      \end{minipage}
      \\
   \end{center}
   \today 
   \vspace{0.6cm}
\end{@twocolumnfalse}
]

\renewcommand*\rmdefault{bch}\normalfont\upshape
\rmfamily
\section*{}
\vspace{-1cm}

\footnotetext{\emph{$^{a}$ Center for Free-Electron Laser Science, Deutsches Elektronen-Synchrotron
      DESY, 22607 Hamburg, Germany}}%
\footnotetext{\emph{$^{b}$ Center for Ultrafast Imaging, Universität Hamburg, 22761 Hamburg,
      Germany}}%
\footnotetext{\emph{$^{c}$ Department of Physics, Universität Hamburg, 22761 Hamburg, Germany}}%
\footnotetext{\emph{$^{d}$ Deutsches Elektronen-Synchrotron DESY, 22607 Hamburg, Germany}}%
\footnotetext{\emph{$^{e}$ Elettra-Sincrotrone Trieste S.C.p.A., 34149, Basovizza, Italy}}
\footnotetext{\emph{$^{f}$ Department of Chemistry, Universität Hamburg, 20146 Hamburg, Germany}}%
\footnotetext{\emph{$^{g}$ European XFEL GmbH, 22869 Schenefeld, Germany}}%
\footnotetext{\emph{$^{h}$ J.R. Macdonald Laboratory, Department of Physics, Kansas State
      University, Manhattan, KS 66506, USA}}%
\footnotetext{\emph{$^{\ast}$sebastian.trippel@cfel.de;
      https://www.controlled-molecule-imaging.org}}

\section{Introduction}
\label{sec:introduction}
Indole, the chromophore of the essential amino acid tryptophan, is an ubiquitous part of peptides
and proteins. It is the strongest near ultraviolet (UV) absorber in these biological molecules and,
for a detailed understanding of the photostability and radiation damage of these biological samples,
it is highly relevant to disentangle indole's intrinsic photophysics, \eg, its various excitation,
relaxation, and fragmentation pathways following electronic excitation. Indole was extensively
studied using microwave~\cite{Suenram:JMolSpec127:472, Caminati:JMolStruct240:253} and optical
spectroscopy~\cite{Philips:JCP85:1327, Berden:JCP103:9596, Brand:PCCP12:4968:2010,
   Kuepper:PCCP12:4980, Korter:JPCA102:7211, Kang:JCP122:174301, Hager:JPC87:2121,
   Short:JCP108:10189}, including vibrationally~\cite{Hager:JPC87:2121, Short:JCP108:10189} and
rotationally resolved~\cite{Philips:JCP85:1327, Berden:JCP103:9596, Brand:PCCP12:4968:2010,
   Kuepper:PCCP12:4980, Korter:JPCA102:7211, Kang:JCP122:174301} electronic spectroscopy, and also
using time-resolved ion and photoelectron spectroscopy~\cite{Montero:JPCA116:2968,
   Livingstone:JCP135:194307, Godfrey:PCCP17:25197}. Here, we extend these studies to the
investigation of the photophysics and photofragmentation dynamics of indole following soft x-ray
absorption.

Fragmentation studies of isolated gas-phase molecules and clusters allow to extract molecular
properties, such as the geometric structure~\cite{Stapelfeldt:PRA58:426, Pitzer:Science341:1096}.
Therefore, they provide a link between the laboratory frame and the molecular frame that allows to
investigate wave packet dynamics on complex potential energy surfaces through molecular-frame
dependent observables such as, for instance, molecular-frame angle-resolved photoelectron
spectroscopy (MF-ARPES)~\cite{PopovaGorelova:PRA94:013412, Boll:PRA88:061402}. Furthermore,
fundamental relaxation processes like Auger decay, interatomic (intermolecular) Coulombic
decay~\cite{Cederbaum:PRL79:4778, Jahnke:JPB48:082001}, or electron-transfer mediated decay
(ETMD)~\cite{Zobeley:JCP115:5076} can be investigated upon x-ray ionization, and can be employed
as observables to study molecular dynamics. In order to understand the complete fragmentation and
charge rearrangement dynamics of molecules and small compound systems such as clusters, coincidence
measurements can be highly advantageous~\cite{Ullrich:RPP66:1463}. Various techniques were developed
during the last years~\cite{Arion:JESRP200:222, Morin:JESRP93:49}, which include photoion-photoion
coincidence (PIPICO), photoelectron-photoion-photoion coincidence (PEPIPICO), or Auger-electron
photoion-photoion coincidence (AEPIPICO) measurements~\cite{Sugishima:JCP131:114309, Boll:FD171:57,
   Wolter:Science354:308, Ablikim:PCCP19:13419, Erk:JPB46:164031, Bomme:RSI84:103104,
   Kukk:PRA91:043417, Levola:PRA92:063409, Ha:PRA84:033419, Ha:JPCA118:1374}. Such coincidence
measurements can, at least for simple molecules, be used to study molecular-frame (MF) properties by
reconstructing the molecular orientation from the measured three-dimensional (3D) velocity
distributions of all charged fragments, which is the recoil-frame (RF) of the molecule.
   The connection between the RF and the MF requires unique molecular fragments, \eg, ``marker
   atoms'', and \emph{prior knowledge} about the directionality of the fragmentation to determine
   the orientation of the molecule within the RF. Studies in the RF include recoil-frame
angle-resolved photoelectron spectra (RF-ARPES)~\cite{Shigemasa:PRL74:359, Bomme:RSI84:103104,
   Toffoli:JCP126:054307, Dowek:EPJST169:85, Sann:PRL117:243002, Guillemin:NATCOMM6:6166}, which
allow to image molecular orbitals and their temporal evolution during
dissociation~\cite{Sann:PRL117:243002}, or to extract structure and molecular dynamics information
by ``diffraction from within''~\cite{Landers:PRL87:013002} type of experiments. For such
experiments, it is highly advantageous to locally ionize the molecule at a specific atom, which can
be achieved by inner-shell ionization \emph{via} extreme ultraviolet radiation, soft x-ray, or x-ray
radiation. Localized ionization provides also access to the local electronic structure and excited
state dynamics~\cite{Sann:PRL117:243002, Wolf:NATCOMM8:29, McFarland:NATCOMM5:4235}, and can be used
to break specific bonds~\cite{Eberhardt:PRL50:1038}.

Here, isolated indole (C$_8$H$_7$N) molecules were ionized by a single (soft) x-ray photon with an
energy of 420~eV, \ie, $\ordsim10$~eV above the nitrogen $1s$ ionization threshold, the N($1s$)
edge. This gives rise to an enhanced localized ionization at the nitrogen atom in the
molecule.\footnote{At a photon energy of 420~eV, the nitrogen atom has the highest atomic cross
   section ($0.6466\cdot10^{-22}\text{~m}^2$) of the molecule's constituents, followed by carbon
   atoms ($0.4327\cdot10^{-22}\text{~m}^2$)~\cite{Elettra:CrossSection:Web}. In total, the indole
   monomer contains eight carbon and one nitrogen atom, leading to a probability of 16~\% that the
   complex is locally ionized out of the the nitrogen $1s$ orbital, assuming that the molecular
   cross sections for the $1s$ orbitals do not differ significantly from the atomic ones, and
   neglecting the contribution from the inner-valence and valence orbitals, which are estimated to
   be on the order of a few percent. The photoabsorption cross section for atomic hydrogen is 3000
   times smaller than for nitrogen and is not taken into account.} Photo- and Auger electrons as
well as the ionic fragments of indole were detected in coincidence in a double-sided velocity map
imaging (VMI) spectrometer (VMIS)~\cite{Eppink:RSI68:3477}. Our work provides the first inner-shell
photoionization study of bare gas-phase indole. It also provides the basis for relaxation and
fragmentation studies of larger indole-containing molecules, \eg, tryptophan, as well as molecular
clusters, such as the investigation of intermolecular interactions in
indole--water~\cite{Trippel:PRA86:033202, Sobolewski:PCCP4:1093} or
indole--ammonia~\cite{Sobolewski:PCCP4:1093}. In fact, the experiment described here was set up such
that the photofragmentation of indole and indole--water clusters could both be measured. Our
findings for the photophysics of indole--water$_1$ clusters are beyond the scope of this manuscript
and will be presented in an upcoming publication~\cite{Kierspel:Dissertation:2016,
   Kierspel:indole-water:inprep}.

\section{Experimental setup}
\begin{figure}
   \centering%
   \includegraphics[width=1\linewidth]{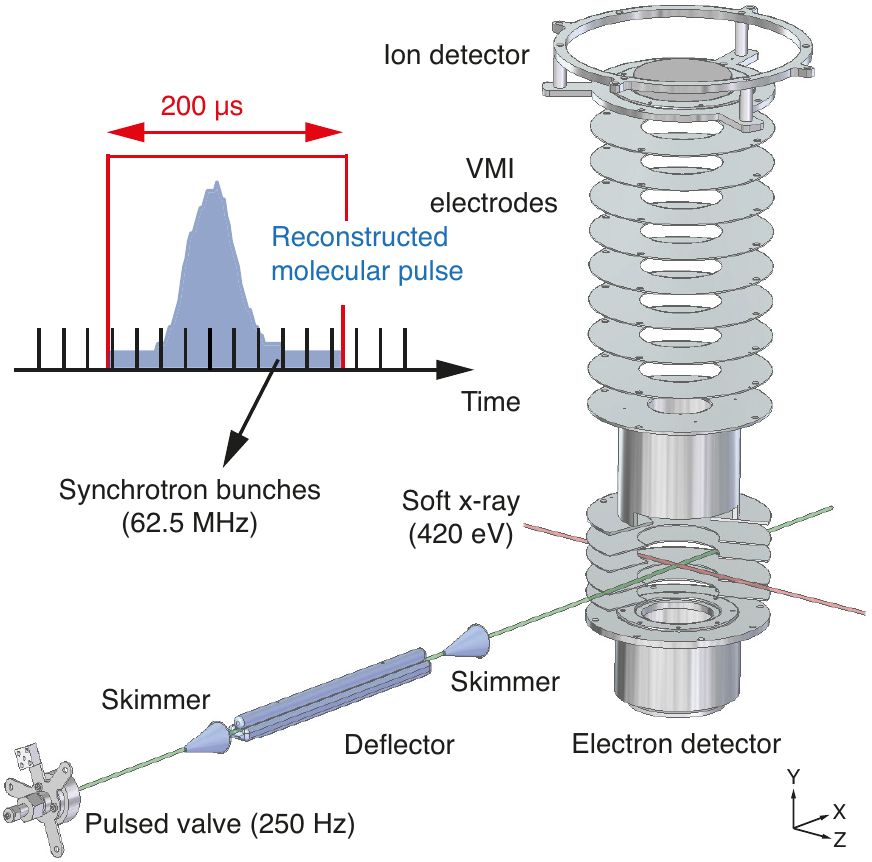}
   \caption{Experimental setup showing the pulsed valve, skimmers, deflector, the double-sided VMIS,
      and the synchrotron beam, which crosses the molecular beam in the center of the
      VMIS~\cite{Bomme:notitle}. A reconstructed molecular pulse is shown in the top left part.
      Schematically indicated is the logical gate (red) synchronized to the molecular beam. Multiple
      synchrotron pulses (black vertical bars) are crossing the molecular beam. Due to the low
      interaction probability of the synchrotron pulses with the molecular beam and background gas,
      only a few events per molecular beam pulse were detected.}
   \label{fig:ind:setup}
\end{figure}

\autoref{fig:ind:setup} shows the experimental setup, including a species-selecting molecular-beam
injector~\cite{Trippel:PRA86:033202, Chang:IRPC34:557}.
\label{sec:ind:setup}
A supersonic expansion of a few mbar of indole seeded in 60~bar of helium was provided by a pulsed
Even-Lavie valve~\cite{Even:JCP112:8068}. The valve was operated at a repetition rate of 250~Hz and
a temperature of \celsius{110}. The deflector was used to spatially separate different species
present in the expansion, including a separation of indole from the helium seed gas.

The molecular beam apparatus was mounted to the CFEL-ASG Multi-Purpose (CAMP)
endstation~\cite{Strueder:NIMA614:483}, which was connected to the Petra~III synchrotron's variable
polarization beamline P04~\cite{Viefhaus:NIMA710:151} (circular polarization $>\;98\%$,
$5\cdot10^{13}$~photons/s, 480 bunches, 16~ns bunch spacing). The molecular beam was crossed by the
420~eV ($\lambda$ = 2.95~nm) synchrotron radiation under an angle of 90~degree inside a double-sided
VMIS~\cite{Bomme:notitle} for simultaneous electron and ion detection. Electrons and ions were
detected with a hexanode (electrons) and quadanode (ions) delay line detector (HEX80 and DLD80,
RoentDek), respectively. For the data presented, however, the hexanode detector had to be operated
as a quadanode due to a defect third delay-line layer. The electronic readout was triggered by the
detection of an electron and was set to an acquisition time of 6~\us, which was long enough to
detect ionic fragments with an atomic mass ($m$)-to-charge ($q$) ratio of up to $\ordsim220$. The
pulse duration of the molecular beam in the interaction region was about 60~\us full width at half
maximum (FWHM), resulting in a duty cycle of $\ordsim1.5~\%$. A logical gate, synchronized to the
arrival time of the molecular beam in the interaction zone, was used to record data in a 200~\us
time window, reducing the absolute number of background events. The overall event rate was on the
order of a few hundred events per second. The inset of \autoref{fig:ind:setup} shows the
reconstructed temporal molecular beam profile plus a constant offset due to background events. The
background events were used as a background correction in, \eg, \autoref{fig:ind:Indole_PIPICO}. In
addition to the reconstructed molecular beam profile vertical black lines are shown, indicating the
pulse structure of the synchrotron.

\section{Coincidence spectra}
\label{sec:ind:coin}
\begin{figure}
   \centering%
   \includegraphics[width=1\linewidth]{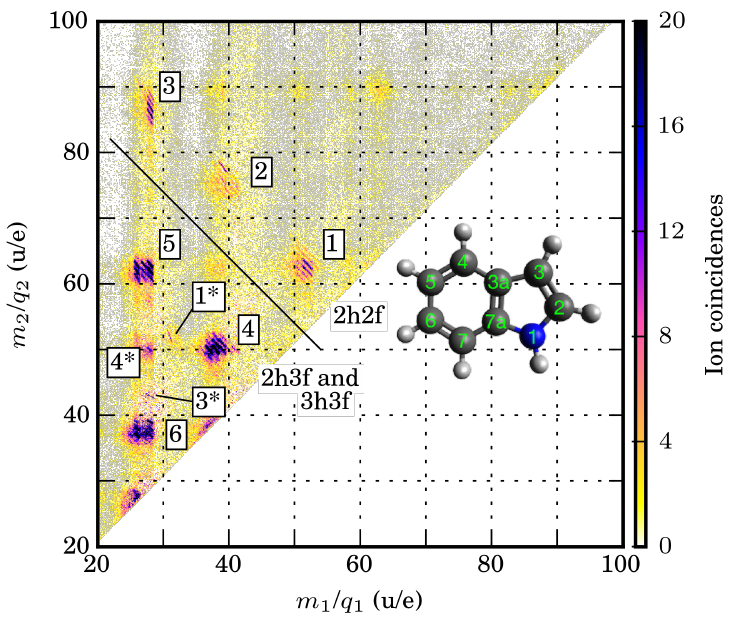}
   \caption{PEPIPICO spectrum of the first two detected ions of indole following inner-shell
      ionization. The inset shows the molecular structure of indole with atomic labeling following
      the IUPAC recommendations~\cite{Moss:PAC70:143}. The solid black line is visually separating
      the 2h2f regions from the other regions.}
   \label{fig:ind:Indole_PIPICO}
\end{figure}

The photofragmentation of indole upon single-photon inner-shell ionization from the nitrogen and
carbon $1s$ orbitals was investigated \emph{via} a coincidence measurement between the emitted
electrons and the corresponding ionic fragments. A background subtracted PEPIPICO
spectrum~\cite{Eland:JESRP41:297, Frasinski:JPB19:L819} of indole is shown in
\autoref{fig:ind:Indole_PIPICO} as a function of the atomic mass-to-charge $m/q$ ratio of the first
and second detected ion, $m_1$/$q_1$ and $m_2$/$q_2$, respectively.
\begin{table*}
   \caption{Overview of the identified ion-fragmentation channels extracted from the ion coincidence
      spectrum shown in \autoref{fig:ind:Indole_PIPICO}. The indices $i$ and $j$ in the formulas
      show the number of hydrogen-atom or proton losses that resulted in separate lines with a
      spacing of $m/q=1~\text{u}/e$ within a given island. Regions 4--6, and 4* consist of three
      heavy neutral/ionic fragments, with numerous different possibilities for hydrogen-atom or
      proton losses, which are thus not listed explicitly.}
   \centering \setlength{\extrarowheight}{4pt}
   \begin{tabular}{@{}lclcccc@{}}
     \hline\hline
     Region & Fragmentation type & Fragmentation channel & mass sum (u) & $i$ & $j$ \\
     \hline
     1 & 2h2f & $\begin{cases}\mathrm{C}_4\mathrm{H}_{4-i}^++\mathrm{C}_4\mathrm{NH}_{3-j}^+ \\ \mathrm{C_3NH}_{3-i}^++\mathrm{C_5H}_{4-j}^+\end{cases}$
            & 112--117 & \begin{tabular}{c}0--1 \\ 0--2\end{tabular} & \begin{tabular}{c}0--3 \\ 0--3\end{tabular} \\
     1* & 3h2f & $\left\{\begin{tabular}{@{\ }l@{}}$\mathrm{C}_4\mathrm{H}_4^++\mathrm{C}_4\mathrm{N}^{++}$ \\
           $\mathrm{C}_3\mathrm{NH}_2^++\mathrm{C}_5\mathrm{H}_2^{++}$\end{tabular}\right.$&  $114$ & 0 & 0\\
     2 & 2h2f  &$\left\{\begin{tabular}{@{\ }l@{}}
           $\mathrm{C}_{2}\mathrm{NH}_{3-i}^++\mathrm{C}_6\mathrm{H}_{4-j}^+$ \\ $\mathrm{C}_3\mathrm{H}_{3-i}^+$ +
           $\mathrm{C}_5\mathrm{NH}_{4-j}^+$
           \end{tabular}\right.$& 112--117 & \begin{tabular}{@{}c@{}}0--3\\0--1\end{tabular}
            & \begin{tabular}{@{}c@{}}0--2 \\0--4\end{tabular}\\
     3 & 2h2f & \quad $\mathrm{CNH}_{2}^++\mathrm{C}_7\mathrm{H}_{5-i}^+$ & 113--117 & 0--4 & 0\\
     3*& 3h2b & \quad $\mathrm{CNH_{2}^++C_7H_{2}^{++}}$ &  $114$ & 0 & 0 \\
     4 & 2h3f / 3h3f & $\left\{\begin{tabular}{@{\ }l@{}}$\mathrm{C_3H_3^++(C_3NH_2^+~or~C_4H_4^+)}$\\
     $\mathrm{C_2NH^++C_4H_4^+}$\end{tabular}\right.$ & 86--91\\
     4* & 2h3f / 3h3f &$\left\{\begin{tabular}{@{\ }l@{}}$\mathrm{C_2H_2^++(C_3NH_2^+~or~C_4H_4^+)}$ \\
           $\mathrm{CNH_2^++C_4H_4^+}$\end{tabular}\right.$& 75--79 &  & \\
     5 & 2h3f / 3h3f & $\left\{\begin{tabular}{@{\ }l@{}}$\mathrm{(C_2H_2^+~or~CNH_2^+)+C_5H_3^+}$ \\
           $\mathrm{C_2H_2^++C_4NH^+}$\end{tabular}\right.$& 87--91 &  &  \\
     6 & 2h3f / 3h3f / ... & $\left\{\begin{tabular}{@{\ }l@{}}$\mathrm{(C_2H_2^+~or~CNH^+)+C_3H_3^+}$ \\
           $\mathrm{C_2H_2^++C_2NH^+}$\end{tabular}\right.$& 61--67 &  &  \\
     \hline\hline
   \end{tabular}
   \label{tab:pipico}
\end{table*}
The molecular structure of indole is shown in the inset of \autoref{fig:ind:Indole_PIPICO}. The
PEPIPICO map allows to disentangle different fragmentation channels of indole in the case of at
least two detected ionic fragments. Nine principal coincidence regions are observed, which are
labeled 1--6, 1$^*$, 3$^*$, and 4$^*$. A detailed list of the identified fragmentation channels is
given in \autoref{tab:pipico}. The sum of the masses of the fragments in regions 1--3 is equal to
the mass of indole, neglecting the loss of hydrogen/protons. Therefore, these fragmentation channels
correspond to the generation of two heavy ionic fragments, which are called in the following a
two-hole two-fragment (2h2f) fragmentation channel. They are visually separated from the other
channels in \autoref{fig:ind:Indole_PIPICO} by the solid black line. Coincidence regions 4--6, and
4* are due to fragmentation into three or more fragments, \ie, the total masses of the first two
detected ions corresponding to a single event do not add up to the mass of the indole monomer. The
missing fragments can be neutral or ionic and the corresponding channels are labeled two-hole
three-fragment (2h3f) and three-hole three-fragment (3h3f), respectively. Due to a limited detection
efficiency, the 3h3f fragments can split into different coincidence regions as, for example, the
regions 4 and 4*. Both regions have the same 'heavy' second detected ion, \ie,
$\mathrm{C_3NH_2^+~or~C_4H_4^+}$, but alternating 'lighter' fragments for the first detected ion. If
only the 'lighter' fragments are detected, or if all ions are detected, this fragmentation channel
is, in the used representation, part of region 6. Regions 1*, and 3* have molecular fragments with
the same masses as regions 1, and 3, but with different charge distribution, \ie, they contain both,
singly and doubly charged ionic fragments and are labeled therefore as three-hole two-fragment
(3h2f) channels.

If not stated otherwise, the losses of hydrogens or protons will not be considered, and are not
included in the labeling of the different fragmentation channels. Further, 2h2f and 2h3f
fragmentation channels are quantified such that they show strong axial recoil, as described
in~\autoref{sec:frag:dyn}. In contrast, the majority of ions detected in 3h3f fragmentation
channels do not show a strong axial recoil. Therefore, if not all ions are detected in a 3h3f
fragmentation channel, these channels are distinguished from 2h2f or 2h3f by their axial recoil.
Furthermore, due to the stronger Coulomb repulsion between three ionic fragments, the kinetic energy
of the 3h3f fragments gives a hint toward these fragmentation channels.

Taking this assumptions into account and assuming an ion detection efficiency $\ordsim40$~\%,
the branching ratios between the main regions of the PEPIPICO
spectrum can be estimated to 27~\%, 51~\%, and 22~\% for 2h2f, 2h3f and 3h2f/3h3f, respectively. The
detection efficiency of the electrons is neglected, leading to an overestimation of the contribution
of 3h2f and 3h3f fragmentation channels. Independent of the electron detection efficiency, the
majority of indole molecules is thus fragmenting into three heavy fragments.

If proton and hydrogen transfer processes are neglected, PEPIPICO region 3 and 3* are the only
PEPIPICO regions for which the ionic fragments can be uniquely assigned, \ie,
$\mathrm{CNH_{2}}+\mathrm{C_7H_{5-i}}$ corresponding to the atoms (1, 2) and (3, 3a, 4, 5, 6, 7,
7a); see the notation in the inset of~\autoref{fig:ind:Indole_PIPICO}. In contrast, PEPIPICO region
1 and 2 consist of a superposition of two fragmentation channels, which can additionally consist of
non-unique fragmentation combinations of the indole molecule. Consider, for example, the
fragmentation $\mathrm{C_3NH_{3-i}}+\mathrm{C_5H_{4-j}}$ of PEPIPICO region~1. The possible atomic
combinations for $\mathrm{C_3NH_{3-i}}$ are (1,2,3,3a), (1,2,3,7a), (1,2,7,7a), or (1,6,7,7a). In
the case of 2h3f and 3h3f fragmentation channels (regions 4--6) the possible combination of ionic
fragments is further increased, resulting in an even lower probability to uniquely assigning the
fragments. Exceptions are some single coincidence lines within a coincidence region, such as
$\mathrm{C_4H_4}+\mathrm{C_4NH_3}$ (PEPIPICO region 1) whose mass sum is equivalent to the mass of
the indole molecule, \ie, including the mass of all hydrogens.

\section{Fragmentation dynamics}
\label{sec:frag:dyn}
\begin{figure}[t]
   \centering
   \includegraphics[width=1\linewidth]{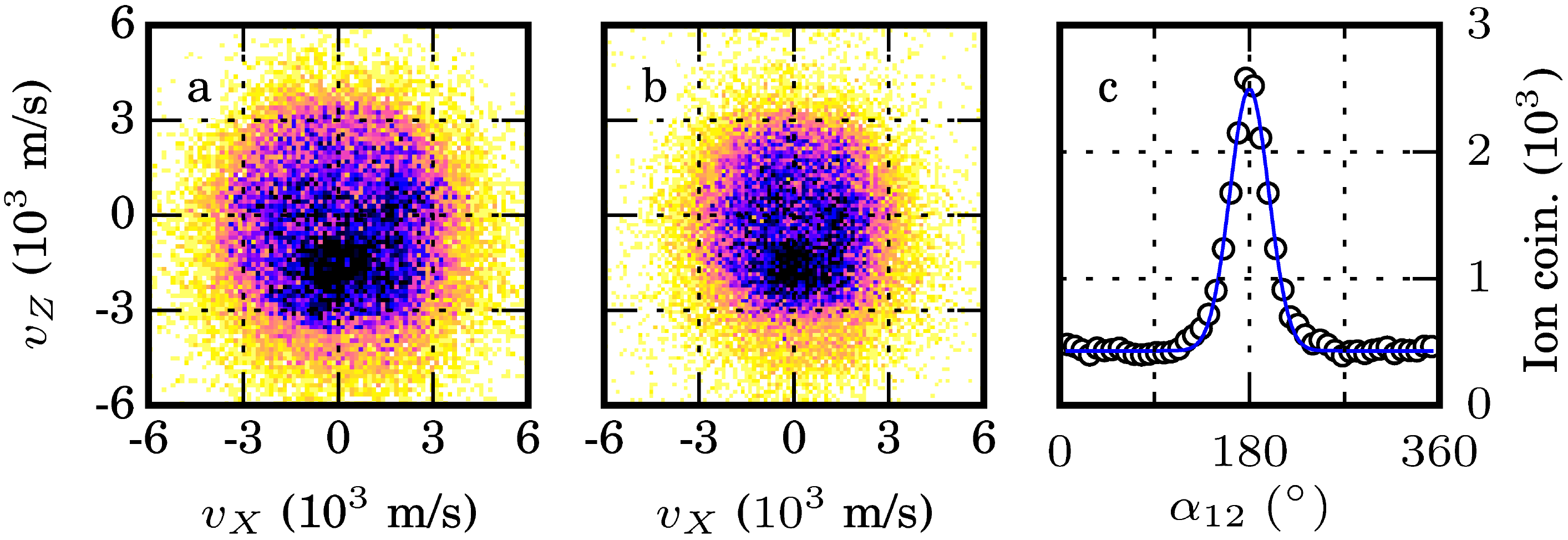}
   \caption{VMI images of the a) first and b) second detected ion contributing to the 2h3f
      fragmentation channel of coincident region 4. c) Histogram of the angle between the first and
      the second ion with a Gaussian fit indicated by the blue line.}
   \label{fig:ind:vmi_2hn}
\end{figure}
The VMIS is used to measure the projected velocity vectors of the ionic fragments.
\subautoref{fig:ind:vmi_2hn}{a and b} show the VMI images for the first and second detected ion in
the coincidence region 4. The corresponding fragments are $\mathrm{C_3H_3^+}$ and
$\mathrm{(C_3NH_2^+~\text{or}~C_4H_4^+)}$ or $\mathrm{C_2NH^+~\text{and}~C_4H_4^+}$; the color scale
is the same as \autoref{fig:ind:Indole_PIPICO}. The velocity of the VMI was calibrated by the
helium--photoelectron recoil for different photon energies ranging from 310 to 420 eV. The first
detected ions show a slightly higher velocity compared to the second detected ions, which is
explained by their smaller mass and the momentum conservation of the fragmenting particles. The
increased number of counts visible in the VMI images at $v_X=0$ and $v_Z\approx-2\cdot10^3$~m/s is
due to background from the carrier gas, which is falsely detected at that corresponding TOF window
and does not obey momentum conservation \footnote{These events might be due to a subsequent pulse of
   the synchrotron radiation ionizing a second particle in the molecular beam within the 6~\us
   acquisition time window (\subautoref{fig:ind:setup}), which has a small but finite probability.
   Helium contributes strongest to the signal from the molecular beam and is, therefore, the main
   background signal.}. A histogram of the angular relationship between the first and second
detected ions is shown in \subautoref{fig:ind:vmi_2hn}{c}. The angle $\alpha_{12}$ is defined as
counter-clockwise rotation about $Z$ starting from the 2D velocity vector of the first detected ion.
The blue line shows a Gaussian fit centered at an recoil angle of $\alpha_{12}=\degree{180}$ with a
standard deviation (SD) of the recoil angle of $\sigma_{\alpha_{12}}=\degree{18.4}$. This strong
axial recoil between ions in this channel is only observed for a 2h3f fragmentation process
(\emph{vide infra}). This is in agreement with the expected fast fragmentation of the molecule due
to Coulomb explosion subsequent to inner-shell ionization, and the momentum conservation between the
ionic fragments. $\sigma_{\alpha_{12}}$ depends on the fragmentation channel, and is
$\sigma_{\alpha_{12}}=\degree{12.7}$ for the 2h2f fragmentation channels, and
$\sigma_{\alpha_{12}}=\degree{9.8}$ and $\sigma_{\alpha_{12}}=\degree{9.5}$ for the 1* and 3*
fragmentation channel, which were assigned to a 3h2f fragmentation channels. These channels show a
stronger confinement in the recoil-frame (RF) because they experience a stronger Coulomb repulsion,
which leads to an RF that is more dominated by Coulomb repulsion. In contrast, in a 2h3f
fragmentation channel the momentum of the Coulomb repulsion is more in competition with the momentum
taken up by the heavy neutral fragment, resulting in a less-confined axial recoil.

The angular variations $\sigma_{\alpha_{12}}$ in the recoil-frame can be expressed as a degree of
(post-)orientation or alignment in the RF, which is $\oricosta\approx0.98$, 0.99, and 0.95, or
$\costa=0.95$, 0.97, and 0.91, for the 2h2f, 3h2f, and 2h3f fragmentation channels, respectively.
The angular confinement, \ie, the alignment, is comparable to the best laser alignment
experiments~\cite{Holmegaard:PRL102:023001} whereas the directionality, \ie, the orientation, is
significantly better~\cite{Holmegaard:PRL102:023001, Trippel:PRL114:103003}. Thus, in the case of
the planar indole molecule, these RF determinations allow for RF-ARPES of the individual ion
fragmentation channels, albeit that the actual angular-resolution quality of the ARPES depend on the
specific fragmentation channel.

The deviation in $\sigma_{\alpha_{12}}$ between the 2h2f and 2h3f can be used to estimate the
velocity of the neutral fragment. An explicit assignment of the neutral fragments of PEPIPICO region
4 and 5 is not possible since the neutral fragments cannot be detected. From the tight momentum
conservation we infer, however, that the bonds between the neutral and the ionic fragments are
broken instantaneously on the timescale of the fragmentation process. In addition, we assume that
the missing masses are intact fragments due to the following reasons: First, the ionic fragments
dominantly stay intact in the case of a 3h3f fragmentation. Second, there is no dominant PEPIPICO
region where only a single carbon is missing. Then, in the case of coincidence region 4 a mean
velocity of 500~m/s can be assigned to a neutral fragment with a mean mass of 27~u.

\begin{figure}[t]
   \centering%
   \includegraphics[width=1\linewidth]{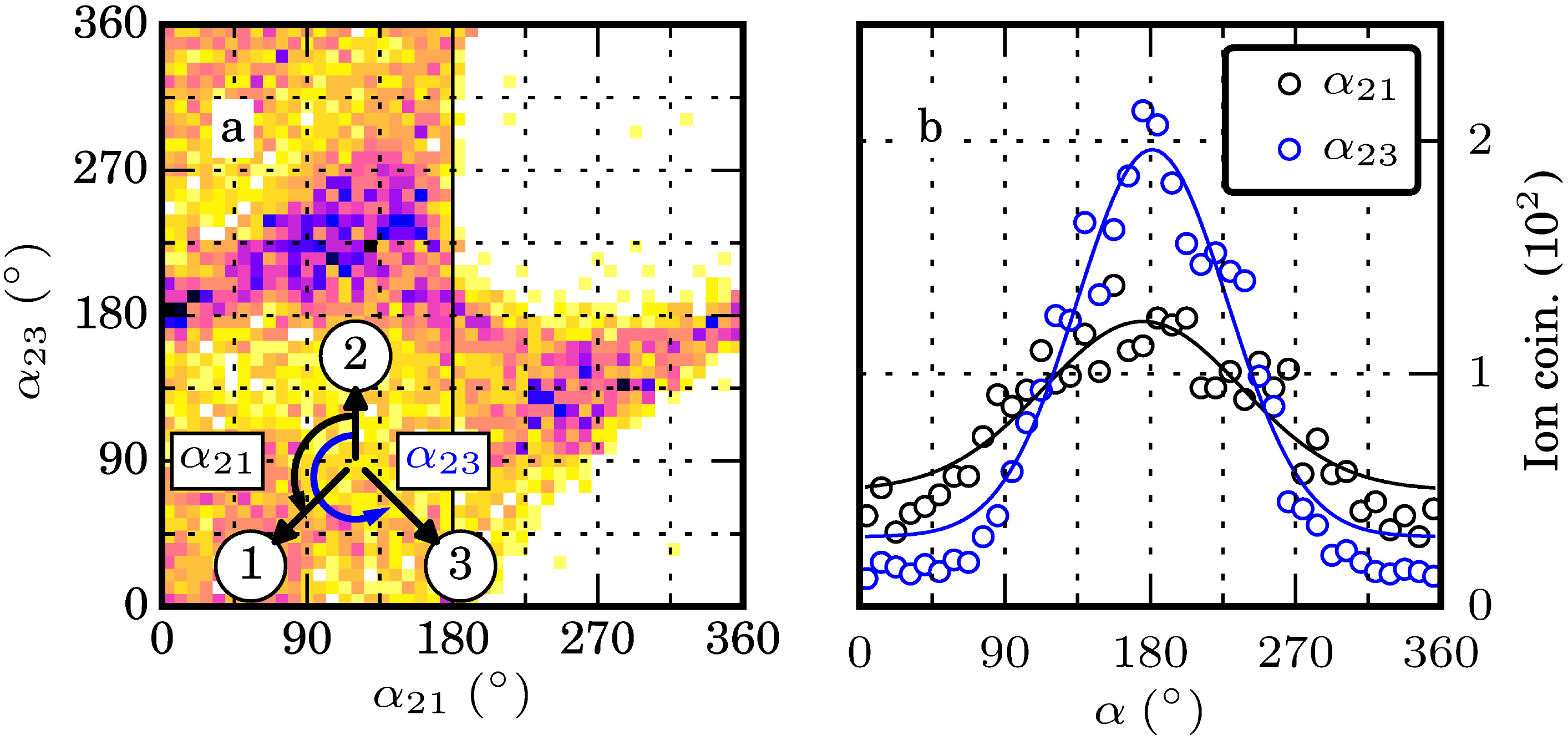}
   \caption{a) Angular relationship between the ions of the 3h3f fragmentation of
      \subautoref{fig:ind:vmi_2hn}. In the right half, only ions that obey momentum conservation are
      shown. The definition of the angle is indicated by the inset in the top right corner.
      $\alpha_\mathrm{21}$ is the angle between the second and first-, $\alpha_\mathrm{23}$ the
      angle between the second and third detected ionic fragment. b) Histograms of the angular
      relationship between the ions of a).}
   \label{fig:ind:vmi_3h}
\end{figure}
\subautoref{fig:ind:vmi_3h}{a} shows the angular correlation between the ions of a 3h3f
fragmentation channel; the second and third detected ions have the same masses as the ions shown in
\autoref{fig:ind:vmi_2hn}, \ie, they correspond to the fragments $\mathrm{C_3H_3^+}$ and
$\mathrm{(C_3NH_2^+~or~C_4H_4^+)}$, or $\mathrm{C_2NH^+~and~C_4H_4^+}$. The first detected ions were
previously neutral and are assigned to the ionic fragments $\mathrm{C_2H_2^+~or~CNH^+}$. The two
dimensional histogram shows the angles $\alpha_{23}$ and $\alpha_{21}$ between the 2D velocity
vector of the second-third and second-first ion pairs. The definition of the angles with respect to
the fragments is visualized by the inset in the top right corner of \autoref{fig:ind:vmi_3h}. The
angular relationship between these pairs of fragments shows an hourglass-like structure, rotated by
approximately \degree{45}. Coincidences outside that structure are due to ions, which do not fulfill
momentum conservation. This is illustrated by right part of the same histogram, where only ion
combinations are shown that do fulfill momentum conservation to a high degree
($<60~\text{u}\cdot117~\text{km/s}$). \subautoref{fig:ind:vmi_3h}{b} shows the histogram of the
angles $\alpha_{21}$ and $\alpha_{23}$ for ion pairs that obey momentum conservation, and allows
therefore for a better comparison of the recoil angle between the 2h3f and 3h3f. These channels have
an SD of $\sigma_{\alpha_{21}}=\degree{70.3}$, and $\sigma_{\alpha_{23}}=\degree{50.7}$, which is a
significantly worse axial recoil compared to the one given in \autoref{fig:ind:vmi_2hn} for a 2h3f
fragmentation channel, and allows therefore to discriminate between both fragmentation channels.
This fixed angular relationship between three heavy ionic fragments demonstrates the possibility to
reconstruct the three dimensional orientation of the molecule in the laboratory frame provided that
the directionality of the moving fragments in the molecular frame are known. Due to the strict
planarity of the indole molecule and the immediate Coulomb explosion, the plane of the molecule can
be assigned to the recoil plane defined by the three ionic fragments. However, the orientation
within the symmetry plane is practically undefined.

\section{Angle-resolved photoelectron spectra}
\label{sec:ind:photoelectron-distributions}
\begin{figure}[t]
   \centering
   \includegraphics[width=1\linewidth]{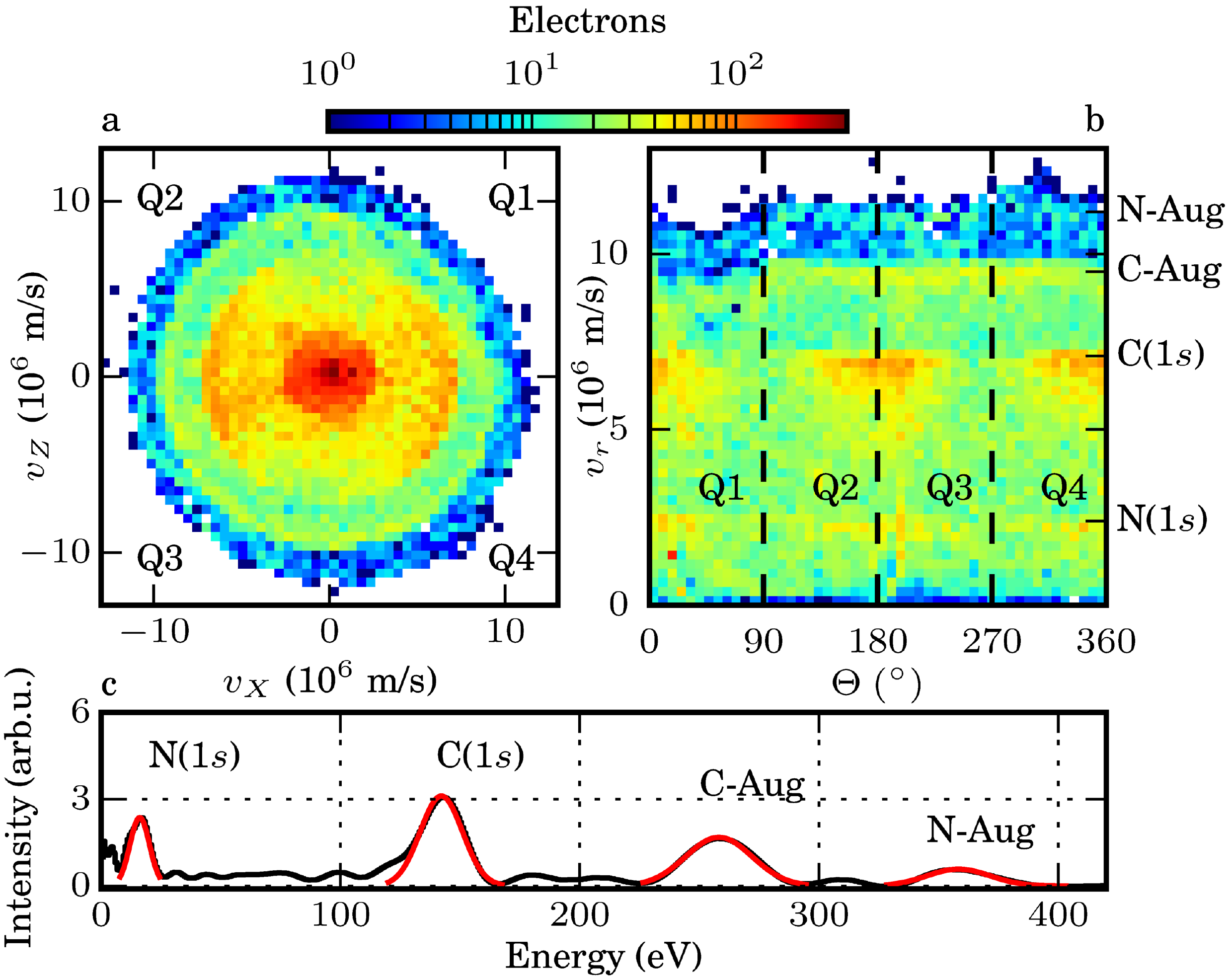}
   \caption{Photoelectron VMI image of indole in cartesian (a) and polar (b) coordinate systems.
      Q1--Q4 indicate the four different quadrants of the VMI image. (c) Photoelectron energy
      spectrum obtained from the inverse-Abel-transformed data of Q2 and Q3 in black. The red
      curves show Gaussian fits to the assigned electron peaks.}
   \label{fig:ind:IndPhotElec}
\end{figure}
\subautoref{fig:ind:IndPhotElec}{a and b} show the electron velocity map in a cartesian and a polar
coordinate system, respectively. The photoelectron VMI has been calibrated by photoelectrons
originating from single-photon ionization of atomic helium and neon, at photon energies between 310
and 980 eV. The labels Q1--Q4 correspond to the four different quadrants of the VMI image; $v_X$ and
$v_Z$ correspond to the electrons velocity component in the laboratory frame, and $v_r$ and $\theta$
are the radial and angular coordinate in the polar coordinate system. The electrons were detected in
coincidence with PEPIPICO regions 1--5, 1* and 3*, with a background correction applied by accepting
only events within $2\sigma$ of the recoil angle of the ions (\autoref{fig:ind:vmi_2hn}). The 3h3f
fragmentation channels of indole have been considered if three ions were detected, if the second and
third detected ion were falling into the coincidence regions 4, and 5, and if the ions fulfilled
momentum conservation (\subautoref{fig:ind:vmi_3h}{b}). Region 6 was not used due to a high number
of background ions detected in this coincidence region. The electron VMI images of indole show four
distinct electron velocities at $2.4$, $7.1$, $9.5$, and $11.2\cdot10^6$~m/s, which correspond to
electron energies of 16, 143, 258, and 358~eV. The additional slow electrons visible in the center
of the VMI image are assigned to background and shake-off electrons from the molecule. The electron
energy spectrum, shown in the bottom graph of \autoref{fig:ind:IndPhotElec}, was obtained by an
inverse Abel transformation based on the BASEX algorithm~\cite{Dribinski:RSI73:2634} of the second
and third quadrant of the electron-VMI image. Quadrants one and four were not used, to avoid the
influence of the VMI distortions in these quadrants, which are visible for velocities grater than
$\ordsim8\cdot10^6$~m/s, and attributed partially to the non-working layer of the hexanode DLD,
possible influence of an magnetic field, or a non well-centered interaction region in the VMI.
Considering atomic electron binding energies, the nitrogen and carbon $1s$ photoelectron energies
would be expected at 10.1 and 135.8~eV~ \cite{Thompson:Xraydata2009}, respectively. In pyrrole
(C$_4$H$_5$N), which corresponds to the five-membered-ring part of indole, the binding energies are
chemically shifted and would correspond to photoelectron energies of 14 and 130~eV for nitrogen and
carbon $1s$, respectively~ \cite{Chambers:JCP67:2596}. This is a deviation of less than 5~\% between
the $1s$ binding energies in pyrrole and indole, which is within the systematic error of our
measurement. The observed C KVV-Auger-electron energies agree with the experimentally
observed lines in benzene at 243--267~eV~\cite{Tarantelli:JCP86:2201}. The N
KVV-Auger-electron energies agree with calculated energies of
356--377~eV~\cite{Thompson:AnalChem48:1336}. Fitted Gaussians, shown by the red line in
\subautoref{fig:ind:IndPhotElec}{c}, allow to extract relative intensities of the specific peaks
and, thus, ratios of the electron channels. By comparing inner-shell ionization events, the N($1s$)
and C($1s$) Gaussian fits show a 26.1~\% probability for localized ionization at the nitrogen atom.
A similar probability of 24.8~\% is obtained by comparing the Auger electron ratio. Both numbers are
slightly higher than the expected probability of 16~\% by considering the atomic cross sections of C
and N. We attribute this difference to the specific properties of the selected Coulomb explosion
channels. The SD of the N($1s$) and C($1s$) photolines are $\sigma=4$ and $\sigma=9$~eV,
respectively, which is attributed to the distortions of the VMIS and the low number of electrons of
the VMI image. The chemical-shift variations of the different carbon atoms ($\ordsim2$~eV) and the
bandwidth of the synchrotron radiation (0.4~eV) are negligible. The anisotropy parameters for the
photo- as well as Auger electrons, obtained from the inverse Abel transformation averaged over the
FWHM of the photoelectron line, are $\beta_{\text{N}(1s)}=1.1~(0.1)$,
$\beta_{\text{C}(1s)}=1.7~(0.1)$, $\beta_{\text{C-Auger}}=0.2~(0.1)$, and
$\beta_{\text{N-Auger}}=0.2~(0.1)$. The anisotropy parameter of the Auger electrons is consistent
with the expected isotropic distribution of electrons in the laboratory frame. The anisotropy
parameter for C($1s$) photoelectrons is slightly lower and the anisotropy parameter for N($1s$)
photoelectrons is significantly lower than the one, $\beta=2.0$, expected for ionization out of an
$s$-orbital by circularly polarized radiation. We attribute this lowered asymmetry
parameters to the interaction of photoelectrons with the potential of the
molecule~\cite{Langhoff:JESRP114:23}, but also partly to the non-perfect reconstruction.

\section{Electron-ion fragmentation correlation}
\label{sec:ind:isdf}
\begin{figure}[t]
   \centering
   \includegraphics[width=1\linewidth]{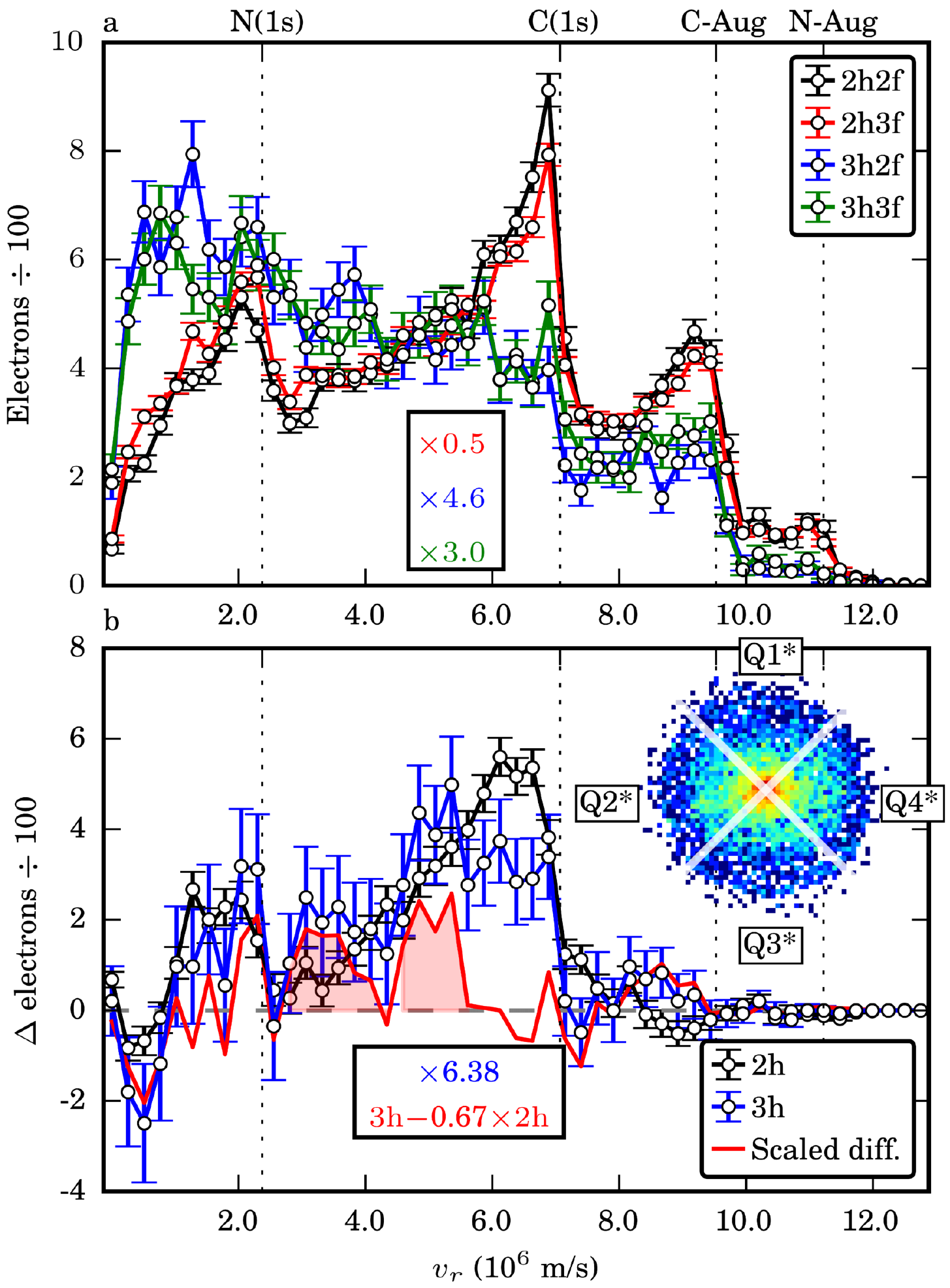}
   \caption{Radial electron-velocity distributions extracted from the electron VMI. The histograms
      are normalized to the same number of electrons; the scaling parameters are given in the inset.
      a) Radial EVD for electrons in coincidence with the ionic fragmentation channels 2h2f, 2h3f,
      3h2f and 3h3f. b) Differential radial plots of the electron VMI retrieved as
      (Q2*+Q4*)-(Q1*+Q3*). The labeling of the quadrants is indicated in the inset, which shows the
      VMI image for electrons detected in coincidence with 3h2f and 3h3f fragmentation channels. }
   \label{fig:ind:pe2h2hn}
\end{figure}
The measured coincidences between electrons and ions allow to extract the individual 2D electron VMI
spectra of the various ionic fragmentation channels. The 2h2f and 2h3f ion fragmentation channels
show a spectrum similar to the one shown in \subautoref{fig:ind:IndPhotElec}{c}. The energy spectrum
of the 3h2f and 3h3f fragmentation channels yielded no clear results due to low statistics.
Therefore, for the 2h2f, 2h3f, 3h2f and 3h3f channels, radial velocities of the electrons 2D VMI
images, \ie, projected electron-velocity distributions (EVD), for the different ionic channels are
compared in the following. This time all quadrants of the electron VMI are taken into account.
The distortions of the VMI (\autoref{fig:ind:IndPhotElec}) in quadrant one and four mainly influenced
the determined energy for the Auger electrons, which do not have a significant influence on the
following discussion.

\subautoref{fig:ind:pe2h2hn}{a} shows histograms of the EVD sorted into the contributions of the
ion-fragmentation channels 2h2f (black), 2h3f (red), 3h2f (blue), and 3h3f (green). The histograms
are normalized to the total number of counts; the multiplication factors are given by the inset, and the
error bars are given as the statistical error. The connecting lines serve to guide the eye. These
electron-velocity distributions clearly group into the two-hole and three-hole channels: The radial
EVD for the 2h2f and 2h3f fragmentation channels (black and red) are very similar. Both show local
maxima of electron counts at velocities assigned to the nitrogen and carbon $1s$ photo- and Auger
electrons. The electrons detected between the maxima are due to the projection of the
three-dimensional electron velocity distribution onto the two-dimensional detector surface. The 2h3f
fragmentation channel has the larger contribution of N($1s$) photoelectrons, whereas the 2h2f
fragmentation channel has larger contributions from C($1s$) photoelectrons and their corresponding
Auger electrons. This indicates a higher probability for a three-fragment break up if indole is
ionized at the nitrogen atom, which can be rationalized by the energy differences between the two
possibilities of ionization: Ionization at the N($1s$) leads to an N KVV-Auger-electron, which
results in a mean energy of 46~eV left in the molecule,\footnote{This
         energy is determined as the difference between the mean photon energy and the mean summed
         electron energies, \ie, the sum of photo- and Auger electron energy.} whereas ionization
at C($1s$) leads to a mean energy of 19~eV. Thus, it seems the larger energy left in the molecule
following N($1s$) ionization than for C($1s$) ionization leads to a stronger fragmentation.

The radial EVD for the three-hole fragmentation channels 3h2f and 3h3f, the blue and green lines in
\subautoref{fig:ind:pe2h2hn}{a}, are also similar. In contrast to the 2h2f and 2h3f radial EVD, the
strongest peak of the spectrum is at electron velocities close to the N($1s$) photoline, and
drops-off continuously toward higher electron velocities, with edges at electron velocities
corresponding to the carbon $1s$ photo- and Auger electrons. This overall shift in the electron
spectrum toward lower photoelectron energies is attributed partially to a tertiary ionization of
indole \emph{via} electron-impact ionization, and also due to satellite peaks of the photo- and
Auger electrons. This is discussed in the second half of the following paragraph based on the
angular anisotropy of the electrons.

To extract an angular anisotropy of the electrons radial distribution, the electron VMI is divided
into the four quadrants Q1*--Q4* as shown in the inset of \subautoref{fig:ind:pe2h2hn}{b}; the
coordinate system is the same as shown in \subautoref{fig:ind:IndPhotElec}{a}, but Q1*--Q4* are
rotated by \degree{45} with respect to Q1--Q4. With $\beta$-parameters of 1.1 and 1.7 for the
nitrogen and carbon $1s$ photoelectrons a larger signal is observed in Q2* and Q4* than in Q1* and
Q3*. For Auger electrons, which typically show no anisotropy, the same averaged number of counts is
expected for all quadrants. The histograms in \subautoref{fig:ind:pe2h2hn}{b} show the radial EVD of
the anisotropy $((Q2^*+Q4^*)-(Q1^*+Q3^*))$ for electrons detected with two and three ionic fragments
in coincidence, \ie, the fragmentation channels 2h2f and 2h3f are jointly labeled 2h (black), and
the fragmentation channels 3h2f and 3h3f are jointly labeled 3h (blue). The error bars depict the
statistical error, the connecting lines serve to guide the eye, and the histograms are normalized to
the number of counts. For the 2h fragmentation channels two distinct maxima are visible at electron
velocities corresponding to the nitrogen and carbon photoelectrons. The anisotropies of the Auger
electrons at $v_\text{r}\gtrsim7\cdot10^6$~m/s are effectively averaged to zero. The negative values
at radial velocities smaller than $1\cdot10^6$~m/s are attributed to non isotropic noise close to
the center of the electron VMI. Comparing the number of electrons assigned to the ionization from
nitrogen/carbon shows a probability of approximately 20~\% for a localized ionization at the
nitrogen atom if the negative values are neglected. This is comparable to the ratio determined from
the overall photoelectron intensities in \autoref{sec:ind:photoelectron-distributions} and, again,
slightly higher than expected from the atomic cross sections. The blue histogram, on the other hand,
shows electrons in coincidence with the 3h fragmentation channels. Here, no clear carbon $1s$
photoelectron line is visible. Instead, an increased number of electrons is detected at velocities
in-between the carbon and nitrogen $1s$ photoelectron energies. Those electron energies can not be
attributed to the earlier determined photo- or Auger electron energies. N($1s$) photoelectrons do
not have enough energy to tertiary ionize indole by electron impact ionization. Also, the
contribution from Auger electrons to triply ionize indole can be excluded in this analysis since
they do not show an anisotropy in the laboratory frame. Therefore, we attribute those electrons to
either inelastically scattered C($1s$) photoelectrons and electrons generated by this inelastic
scattering through electron impact ionization, or to satellite peaks from the C($1s$)
photoelectrons. A closer insight is given by the red line in \subautoref{fig:ind:pe2h2hn}{b}, which
shows a scaled difference between the blue and black spectrum. The scaling was done by a
normalization of the number of electrons at $v_\text{r}=6.8~\cdot10^6$~m/s to subtract the highest
possible contribution from direct photoelectrons. This difference-spectrum shows three main areas:
the contribution of the nitrogen $1s$ photoelectrons and two highlighted red areas, which are
assigned to those inelastic scattered carbon $1s$ photoelectrons, electrons emitted upon impact
ionization, and satellite peaks from the carbon $1s$ photoline. These electrons in the red areas
have a velocity of $v_\text{r}=2.9$--$4.5\cdot10^6$~m/s (24-58 eV) and
$v_\text{r}=4.7$--$5.7\cdot10^6$~m/s (63--92 eV). The number of electrons that correspond to these
two peaks is about the same, and the sum of the mean electron energy of both peaks is 104~eV.

In \subautoref{fig:ind:pe2h2hn}{a}, the C($1s$) Auger- and photoelectrons show a similar behavior,
\ie, the 2h fragmentation channels show a prominent peak, which is absent in the 3h fragmentation
channels. Therefore, we attribute this change in the radial EVD of Auger electrons also to electron
impact ionization or satellite peaks accompanying the Auger electrons.

A quantitative statement about the contribution of the inelastically scattered electrons, electrons
from impact ionization, and satellite electrons to the 3h2f and 3h3f fragmentation channels
could, in principle, be extracted from their anisotropy parameter. This was not possible due to the
low number of detected electrons. Only for C($1s$) photoelectrons a lower limit of 43~\% can be
estimated from \subautoref{fig:ind:pe2h2hn}{b} by counting the number of inelastically
scattered/satellite electrons (red), which are part of the 3h2f and 3h3f channels (blue).

At the given C($1s$) photoelectron energy, the atomic cross section for carbon for electron impact
ionization and elastic scattering of electrons are both in the order of
$200\cdot10^{-22}\text{~m} ^2$~\cite{Kim:PRA66:1279, NIST:ElasticScattering64:2016}. This implies
that elastically-scattered electrons can be detected at comparable signal strengths, \eg, in
photoelectron holography experiments~\cite{Krasniqi:PRA81:033411}. The inelastically-scattered
electrons detected here could be separated by an energy-resolving detection scheme, as demonstrated
here.

\section{Conclusion}
\label{sec:ind:conclusion}
We have performed a detailed photoionization and photofragmentation study of indole upon
single-photon inner-shell ionization at a photon energy of 420~eV. This photon energy was chosen
such that indole could be locally ionized at its nitrogen atom. Ionization from C($1s$) was also
possible and is the dominant ionization process due to the larger number of carbon atoms present in
the molecule. Electrons and ions have been measured in coincidence in a velocity-map-imaging mode to
extract 2D and 3D velocity vectors of the charged particles.

In the ion-coincidence spectrum of indole, \ie, for the events with more than one ionic fragment
observed, indole is fragmenting into two heavy ionic and one neutral fragment in 51~\% of the cases.
These ``heavy'' fragments contain, almost exclusively, two or more heavier atoms; the loss of
hydrogen atoms and protons was also observed, but they were not considered as specific fragments.
Fragmentation channels with only two fragments or with three heavy ionic fragments have also been
observed and showed contributions of 27~\% and 22~\%, respectively. The PEPIPICO spectrum revealed
that the unique assignment of a coincidence region to a carbon atom from a specific position in the
molecule is rather the exception than the rule.

The ion-VMI images could be used to reconstruct the recoil-frame of the molecules. The fragmentation
process was dominated by the Coulomb repulsion of the generated charges. Influence of chemical
effects, \eg, the specific potential-energy surfaces, was observed in the recoil frame of the ions
for the case of a coexisting heavy neutral fragment. Ion-VMI images of this selected 2h3f
fragmentation channel were discussed regarding the velocity of the dissociating neutral fragment,
showing that the bonds between the neutral and ionic fragments must be broken instantaneously on the
timescale of the fragmentation process, \ie, no meta-stable ionic fragments were observed.
Fragmentation channels with three ionic fragments also showed a fixed angular
relationship. This allowed us, for these channels, to directly determine the alignment of the
molecular plane in the laboratory frame. Therefore, the recoil-frame and thus, due to the
symmetry plane of the molecule, the molecular-frame alignment of the molecular plane in the
laboratory frame is uniquely recovered. However, in order to fully reconstruct the
three-dimensional alignment and orientation of the indole molecule, \ie, also the orientation
inside the molecular plane, the direction of the fragments in this plane would have to be known.
This would require elaborate theoretical analysis and is beyond the scope of this paper.

The electron-energy spectrum showed four peaks, which were assigned to photo- and Auger electrons
resulting from element-specific ionization at indole's nitrogen as well as carbon atoms. The
corresponding asymmetry parameters of these peaks were extracted from an inverse Abel
transformation. For the Auger electrons they were isotropic in the laboratory frame, as expected.
For the photoelectrons, deviation from the expected asymmetry parameter for photoelectrons from the
carbon and nitrogen $1s$ orbitals have been observed; where ``expected'' refers to the asymmetry
parameter for a single-photon $1s$ ionization with circularly polarized light. The observed
deviation is partly attributed to the interaction of the photoelectrons with the molecular
potential, partly due to a non-perfect reconstruction of the asymmetry parameters, as well as
deviations due to background signal from slow background and shake-off electrons.

The correlation between ions and electrons showed that different ion fragmentation channels have
different electron spectra, \ie, a relationship between the ionization/excitation process, the
corresponding electronic states, and the fragmentation process, reflecting the specific potential
energy surface. This was shown, for instance, by a comparison of the projected electron energy
spectra for the 2h2f and 2h3f fragmentation channels. In this case it was concluded that inner-shell
ionization at the nitrogen edge leads to a higher probability for indole to break up into three
heavy fragments.

Evidence for secondary electron-impact ionization as well as satellite photoelectrons was observed
in the fragmentation channels where three ionic fragments have been measured. Those channels showed
less pronounced photolines, primarily observed for the C(1$s$) photoelectrons, as well as signals at
electron energies where no photoline is expected. In addition, evidence for satellite peaks of the
Auger electrons and inelastically scattered Auger electrons was presented.

Since the cross sections for the observed inelastic scattering and elastic scattering are comparable
under the experimental conditions, the possibility of photoelectron-holography experiments is
confirmed.

The presented data allowed to record RF-ARPES images of strongly post-oriented indole, albeit that
the relation of RF and MF is unknown beyond the common symmetry plane. Due to the low number of
events per unique fragmentation channels, \ie, fragmentation channels where specific carbon atoms
could be assigned uniquely to the ionic fragment, no statistically significant asymmetries of the
electron distribution in the recoil-frame were observed.

Overall, our results show that the fragmentation channels depend on the different electronic states,
\ie, the chemical potential energy surface, whereas the observed velocities of the fragments are not
strongly dependent of these chemical details.

Our work provides the basis for fragmentation studies of larger molecules as well as molecular
clusters, such as the indole-derivative tryptophan or indole-water clusters.
Comparison of the fragmentation channels and dissociation energies will allow to study the role of
solvents on the photophysics of indole upon site specific x-ray ionization. Furthermore, the
processes observed here provide information on the indole-chromophore-related radiation damage
occurring in coherent diffractive imaging of proteins~\cite{Neutze:Nature406:752, Barty:ARPC64:415}.

\section{Acknowledgments}
We acknowledge Evgeny Savelyev for support with the experiment, and Ludger Inhester
for fruitful discussions about the photofragmentation of indole. Besides DESY, this work has been
supported by the excellence cluster ``The Hamburg Center for Ultrafast Imaging -- Structure,
Dynamics and Control of Matter at the Atomic Scale'' of the Deutsche Forschungsgemeinschaft (CUI,
DFG-EXC1074); by the Helmholtz Association through the Virtual Institute 419 ``Dynamic Pathways in
Multidimensional Landscapes'', the Helmholtz Young Investigators Program (D.R.\ and S.B.), and the
``Initiative and Networking Fund''; by the European Union's Horizon 2020 research and innovation
program under the Marie Skłodowska-Curie Grant Agreement 641789 MEDEA, and by the European Research
Council under the European Union's Seventh Framework Programme (FP7/2007-2013) through the
Consolidator Grant COMOTION (ERC-Küpper-614507). D.R.\ also acknowledges support from the U.S.\
Department of Energy, Office of Science, Basic Energy Sciences, Chemical Sciences, Geosciences, and
Biosciences Division (DE-FG02-86ER13491). S.B.\ also acknowledges support from the Deutsche
Forschungsgemeinschaft (B03/SFB755).

\bibliography{string,cmi}%
\bibliographystyle{rsc}
\balance
\end{document}